\documentstyle[12pt]{article}

\begin{document}
\thispagestyle{empty}
\begin{center}
\LARGE \tt \bf{Riemannian collineations in General Relativity and in Einstein-Cartan cosmology}
\end{center}
\vspace{1cm}
\begin{center} {\large L.C. Garcia de Andrade\footnote{Departamento de
F\'{\i}sica Te\'{o}rica - Instituto de F\'{\i}sica - UERJ
Rua S\~{a}o Fco. Xavier 524, Rio de Janeiro, RJ
Maracan\~{a}, CEP:20550-003 , Brasil.
e-mail.: garcia@dft.if.uerj.br}}
\end{center}
\vspace{1.0cm}
\begin{abstract}
Riemannian vectorial collineations along with current Killing conservation are shown to lead to tensorial collineations for the energy-stress tensor in general relativity and in Einstein-Cartan Weyssenhoff fluid cosmology.
\end{abstract}
\vspace{1.0cm}       
\begin{center}
\end{center}
\newpage
\pagestyle{myheadings}
\markright{\underline{Particle creation on chaotic inflationary models}}
\paragraph*{}
In this note we show that the Riemannian collineations of the energy momentum tensor is obtained when the conserved Killing current is obtained from the beginning.Let us start by defining the vector Killing type current  
\begin{equation}
J^{a}=T^{ab}{\epsilon}_{b} 
\label{1}
\end{equation}
and
By considering the Riemannian covariant derivative operator $D_{a}$ applied on equation (\ref{1}) we obtain 
\begin{equation}
D_{a}J^{a}=(D_{a}T^{ab}){\epsilon}_{b}+T^{ab}D_{a}{\epsilon}_{b}
\label{2}
\end{equation}
In the case ${\epsilon}_{a}$ is a Killing vector the following condition is obeyed
\begin{equation}
D_{({a}}{\epsilon}_{b})=0
\label{3}
\end{equation}
Thus expression (\ref{3}) yields automatically the conserved current since the 
energy-stress tensor in GR is conserved $(D_{a}T^{ab})=0$.Therefore the conserved equation is immeadiatly given by 
\begin{equation}
D_{a}J^{a}=0
\label{4}
\end{equation}
Nevertheless in other theories like Einstein-Cartan the energy-stress is not automatically conserved like in GR,therefore one cannot use the Killing vector current on a trivial way.To include alternative gravity theories other than GR in our scheme we suggest that more general type of Riemannian or even non-Riemannian collineations can be used.Notice that if we assume current conservation from the beginning the expression (\ref{2}) one obtains  
\begin{equation}
(D_{a}T^{ab}){\epsilon}_{b}=-T^{ab}D_{a}{\epsilon}_{b}
\label{5}
\end{equation}
Thus the recurrent spacetime \cite{1} producing the Riemannian vector collineation
\begin{equation}
D_{a}{\epsilon}_{b}= C_{ab}^{c}{\epsilon}_{c}
\label{6}
\end{equation}
Substitution of (\ref{6}) into (\ref{5}) yields
\begin{equation}
(D_{a}T^{ac}){\epsilon}_{c}=-C_{ab}^{c}{\epsilon}_{c}T^{ab}
\label{7}
\end{equation}
As long as the recurrent vector ${\epsilon}$ is linearly independent equation (\ref{7}) yields
\begin{equation}
(D_{a}T^{ac}){\epsilon}_{c}=-C_{ab}^{c}T^{ab}
\label{8}
\end{equation}
Therefore the collineation coefficients C can be determined from this tensorial collineation.We shall now give a simple example in GR although formula (\ref{8}) is equally valid in Einstein-Cartan or other type of alternative gravity.Let us consider the EMT of the type
\begin{equation}
T^{ab}={\rho}u^{a}u^{b}
\label{9}
\end{equation}
of a dust fluid.Thus by applying (\ref{9}) to (\ref{8}) one is able to determine the coefficients $C_{ab}^{c}$.Waldyr AQUI DeIXO essa demonstracao para o Shariff.Acho que em alguns dias faco uma demonstracao no caso de Einstein-Cartan, ok?Um abraco.Luiz Carlos Garcia de Andrade.
\begin{equation}
{\rho}_{r}=6{\pi}G{\sigma}^{2}-[V+\frac{V'}{3H+{\eta}g^{2}{\phi}^{2}}]^{2}
\label{9}
\end{equation}
By equating the RHS of equations (\ref{6}) and (\ref{8}) one obtains a differential equation for the potential $V$ as
\begin{equation}
4V({\phi})=[\frac{V'}{3H+{\eta}g^{2}{\phi}^{2}}]^{2}
\label{10}
\end{equation}
By solving equation (\ref{10}) one obtains the following potential
\begin{equation}
V({\phi})=36H^{2}{\phi}^{2}+6H{\eta}g^{2}{\phi}^{3}+\frac{{\eta}^{2}g^{4}}{4}{\phi}^{4} 
\label{11}
\end{equation}
which is clearly a potential for the chaotic inflation.Throughout the computations we assume the following approximation ${\eta}g^{2}{\phi}<<H$.Finally by substituting expression (\ref{11}) into equation (\ref{9}) one obtains an expression for the spin-torsion density
\begin{equation}
{\sigma}^{2}={\rho}_{r}+4{\phi}+36H^{2}{\phi}^{2}+\frac{{\eta}^{2}g^{2}}{H}(1+\frac{{g}^{2}}{36H}+6\frac{H^{2}}{{\eta}}){\phi}^{3}+\frac{{\eta}^{2}g^{4}}{4}{\phi}^{4}
\label{12}
\end{equation}
which shows that the spin-torsion density possess a kink potential term.A simple expression for the spin-torsion density perturbation in terms of the inflaton potential can be obtained by making use of the quantum fluctuation on the inflaton ${\delta}{\phi}=\frac{H^{2}}{2{\pi}}$ as
\begin{equation}
{\delta}{\sigma}^{2}={\delta}{\rho}_{r}+2\frac{H^{2}}{{\pi}}+\frac{36H^{4}}{{\pi}}{\phi}+\frac{H^{2}{\eta}^{2}g^{2}}{4{\pi}}(1+\frac{{g}^{2}}{36H}+6\frac{H^{2}}{{\eta}}){\phi}^{2}+\frac{H^{2}{\eta}^{2}g^{4}}{2{\pi}}{\phi}^{3}
\label{13}
\end{equation}

A more detailed investigation on the matters disussed here may appear elsewhere.
\section*{Acknowledgments}
\paragraph*{}
I am very much indebt to Professors R.Ramos for enlightening discussions on the subject of this paper.Financial supports from CNPq. (Brazilian Government Agency) and Universidade do Estado do Rio de Janeiro (UERJ) are gratefully acknowledged. 

\end{document}